\documentclass[aps, prl, twocolumn,10pt, superscriptaddress, groupedaddress]{revtex4-1}

\usepackage{amsmath}
\usepackage{amssymb} 
\usepackage{graphicx}
\usepackage{color}
\usepackage{titlesec}
\usepackage[colorlinks, linkcolor = blue, citecolor = blue, filecolor = blue, urlcolor = blue, bookmarks=true, breaklinks=true]{hyperref}

\definecolor{seccolor}{RGB}{25,25,120}
\definecolor{subseccolor}{RGB}{20,90,15}
\definecolor{somtitlecolor}{RGB}{200,35,20}

\titleformat{\section}[hang]	
  {\bfseries\sffamily\Large\color{seccolor}} 
  {\color{seccolor}\thesection} 
  {1ex} 
  {} 
  [] 

\begin{document}

\title{Laser-sub-cycle two-dimensional electron momentum mapping using orthogonal two-color fields}

\author{Li\,Zhang$^{1}$}
\author{Xinhua\,Xie$^1$}
\author{Stefan\,Roither$^1$}
\author{Daniil\,Kartashov$^1$}
\author{YanLan\,Wang$^3$}
\author{ChuanLiang\,Wang$^3$}
\author{Markus\,Sch\"{o}ffler$^1$}
\author{Dror\,Shafir$^{2,5}$}
\author{Paul\,B.\,Corkum$^2$}
\author{Andrius\,Baltu\v{s}ka$^1$}
\author{Igor\,Ivanov$^4$}
\author{Anatoli\,Kheifets$^4$}
\author{XiaoJun\,Liu$^3$}
\author{Andr\'e\,Staudte$^2$}
\email{andre.staudte@nrc-cnrc.gc.ca}
\author{Markus\,Kitzler$^1$}
\email[Corresponding author. ]{markus.kitzler@tuwien.ac.at}

\affiliation{$^1$Photonics Institute, Vienna University of Technology, 1040 Vienna, Austria}
\affiliation{$^2$Joint Laboratory for Attosecond Science of the National Research Council and the University of Ottawa, Ottawa, Ontario, Canada K1A 0R6}
\affiliation{$^3$State Key Laboratory of Magnetic Resonance and Atomic and Molecular Physics, Wuhan Institute of Physics and Mathematics, Chinese Academy of Sciences, Wuhan 430071, China}
\affiliation{$^4$Research School of Physical Sciences, The Australian University, Canberra, ACT 0200, Australia}
\affiliation{$^5$Department of Physics of Complex Systems, Weizmann Institute of Science, Rehovot 76100, Israel}

\begin{abstract}
We study laser-sub-cycle control over electron trajectories concomitantly in space and time using orthogonally polarized two-color laser fields. 
We compare experimental photoelectron spectra of neon recorded by coincidence momentum imaging with photoelectron spectra obtained by semi-classical and numerical solutions of the time-dependent Schr\"odinger equation. We find that a resolution of a quarter optical cycle in the photoelectron trajectories can be achieved.
It is shown that depending on their sub-cycle birth time the trajectories of photoelectrons are affected differently by the ion's Coulomb field.
\end{abstract}


\maketitle


Strong optical fields provide control over the shape of photoelectron trajectories on a sub-optical-cycle time scale. Among the many approaches that employ shaped optical fields for trajectory control, orthogonally polarized two-color (OTC) laser pulses represent a simple yet elegant approach to steer field-ionized electron wave packets not only in time but concomitantly also in space \cite{Kitzler2005, Kitzler2006, Kitzler2007, Kitzler2008, Shafir2009, Zhang2014, Kim2005a, Niikura2010, Niikura2011}.
Here, we study the control over electron trajectories with OTC pulses by comparison of experimental photoelectron momentum spectra to semi-classical and numerical solutions of the time-dependent Schr\"odinger equation (TDSE).
We find that a quarter optical cycle resolution in the trajectory can be achieved. We use the resolution provided by the OTC field to study the influence of the parent ion's Coulomb potential on recolliding and non-recolliding trajectories of field-ionized photoelectrons. We observe a surprisingly strong distortion of non-recolliding trajectories that causes the loss of the expected space-time coupling below a quarter of an optical cycle.

Although experimental photoelectron spectra could be successfully explained by different models that neglect the influence of the ionic field on the emitted electron wave packets, the importance of the Coulomb potential is by now well appreciated and has been demonstrated in many experiments, e.g. \cite{Quan2009, Blaga2009, Huismans2010}, and numerical simulations, e.g. \cite{Brabec1996a, Chelkowski2004, Dimitriou2004}. Inclusion of the Coulomb force into the theoretical description of electron momentum spectra is not straightforward, though \cite{Smirnova2008, Chen2009b}. Likewise,
clearly separating and identifying Coulomb contributions in the experimental electron spectra has remained a challenge with only few successful attempts, e.g. \cite{Shafir2013, Xie2013}.

%
\noindent 
\begin{figure}[tb]
\centering
\includegraphics[width=0.95\columnwidth, angle=0]{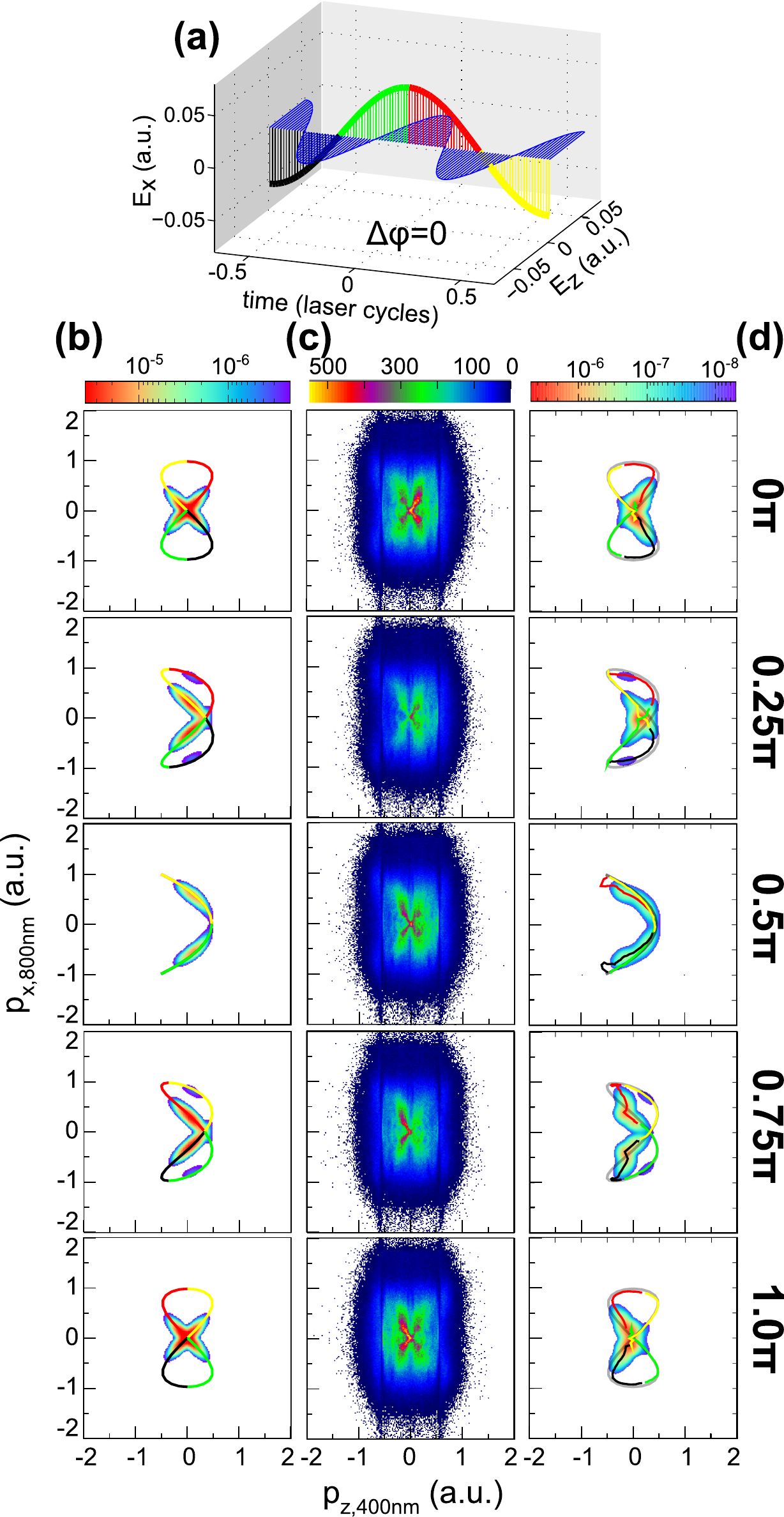}
\caption{(a) Electric fields of the 400\,nm (blue) and 800\,nm pulse. The colors encode quarter cycles of the 800\,nm field. (b) Field-driven electron momentum $\vec{p}=-\vec{A}(t_0)$ for a single cycle of the OTC field in the plane of polarization for various phases $\Delta \varphi$ between fundamental and second harmonic. 
(c) Measured electron momentum distributions correlated to Ne$^+$ in the $xz$-polarization plane of the OTC pulse with $|p_y|<0.1$\,a.u..
(d) Electron momentum spectra calculated the same way as those in (b) but with $\vec{p}(t_0)$ extracted from Fig.\,\ref{fig3}(d), thereby accounting for the parent ion's influence. The momentum $\vec{p}=-\vec{A}(t_0)$ is underlaid in gray. 
}\label{fig1}
\end{figure}

Fig.\,\ref{fig1} illustrates the mapping of emission time to the angle in the laser polarization plane of electron momentum spectra that is established by the OTC pulses. Fig.\,\ref{fig1}(a) shows one cycle of an OTC pulse consisting of an 800\,nm field, linearly polarized along $x$, and a superimposed, orthogonally polarized 400\,nm field along $z$ (blue) of equal peak intensity. The electric field of the OTC pulse is given by
\begin{equation}
\vec{E}(t)=\hat{E}
\left[ 
f_{x}(t)\cos(\omega t)\vec{e}_x+f_{z}(t)\cos(2\omega t+\Delta \varphi)\vec{e}_z 
\right]\label{OTC_field}
\end{equation}
with $\hat{E}$ the peak field strength, $f_{x,z}$ the pulse envelopes along the 800 and 400\,nm direction, respectively, and 
$\Delta \varphi$ the relative phase between the two colors. 
Ignoring the binding potential, wave packets detached by the OTC field within different laser quarter-cycles are observed in different momentum regions in the polarization plane \cite{Kitzler2007, Kitzler2008}.
This is shown in Fig.\,\ref{fig1}(b) for different relative phases by the same color code as the field's quarter-cycles in Fig.\,\ref{fig1}(a).
A change in the phase of the 400\,nm field relative to the 800\,nm field changes the OTC field and therewith the driving force for the trajectories of field ionized electron wave packets on sub-cycle time-scales, which leads to different time-to-momentum mapping for different $\Delta \varphi$ \cite{Kitzler2007, Kitzler2008}.

The mapping is based on the classical relation $\vec{p}=-\vec{A}(t_0)$ obtained by neglecting the influence of the ion on the emitted electrons \cite{Ivanov2005}. Here, $\vec{A}(t)=-\int_{-\infty}^t \vec{E}(t') \text{d}t'$ is the vector potential and $t_0$ the electron birth time. This relation is widely used to interpret experimental electron spectra, often to extract sub-cycle timing information.
In this simplest picture of strong field ionization the graphs in Fig.\,\ref{fig1}(b) delineate the accessible momentum space for photoelectrons in OTC pulses. 
A spectrum of photoelectrons is obtained if also the ionization probability in the OTC field is taken into account. 
Electron momentum spectra predicted by this semi-classical model, referred to as the strong-field classical trajectory (SFCT) model \cite{Xie2013}, are shown in Fig.\,\ref{fig1}(b) in comparison to the classical relation discussed above. 
In the SFCT model, a spectrum is obtained by integration of the classical relation $\vec{p} = -\vec{A}(t_0)$. The spectrum represents the incoherent sum over all possible birth times $t_0$ at which a wave packet is emitted at the origin with a probability determined by the ionization rate \cite{Yudin2001a}. 
To account for the momentum width of the electron wave packets a narrow Gaussian distribution around $\vec{p}=-\vec{A}(t_0)$ is introduced.

Experimentally, OTC pulses were focused into a supersonic gas jet of neon atoms and the three-dimensional (3D) momentum spectra of the resulting singly charged ions and correlated electrons were measured with a cold target recoil ion momentum spectroscopy (COLTRIMS) setup \cite{Ullrich1997}. 
The OTC pulses with a field given by Eq.\,\ref{OTC_field} were generated by combining a 46\,fs, 800\,nm ($\omega$) laser pulse, and a 48\,fs, 400\,nm ($2\omega$) pulse, in a collinear geometry at a rate of 5\,kHz. The peak intensity in either color was $\hat{E}^2 = I_{800\mathrm{nm}} = I_{400\mathrm{nm}} = (1\pm0.1)\times10^{14}$\,W/cm$^2$. 
Temporal overlap of the two pulses was ensured by compensating for the different group velocities of the two colors with calcite plates and a pair of fused silica wedges. The latter were also used to vary $\Delta \varphi$ with a precision of roughly 0.3\,as. Calibration of $\Delta \varphi$ was performed by comparison of the measured and simulated Ne$^+$ yield in the spectral cut-off region, see \cite{suppl_mat} for details.
Further details of the optical and the COLTRIMS setup can be found in Refs.\,\cite{Zhang2014} and \cite{Xie2013}, respectively.

%
\noindent 
\begin{figure*}[tb]
\includegraphics[width=1.95\columnwidth]{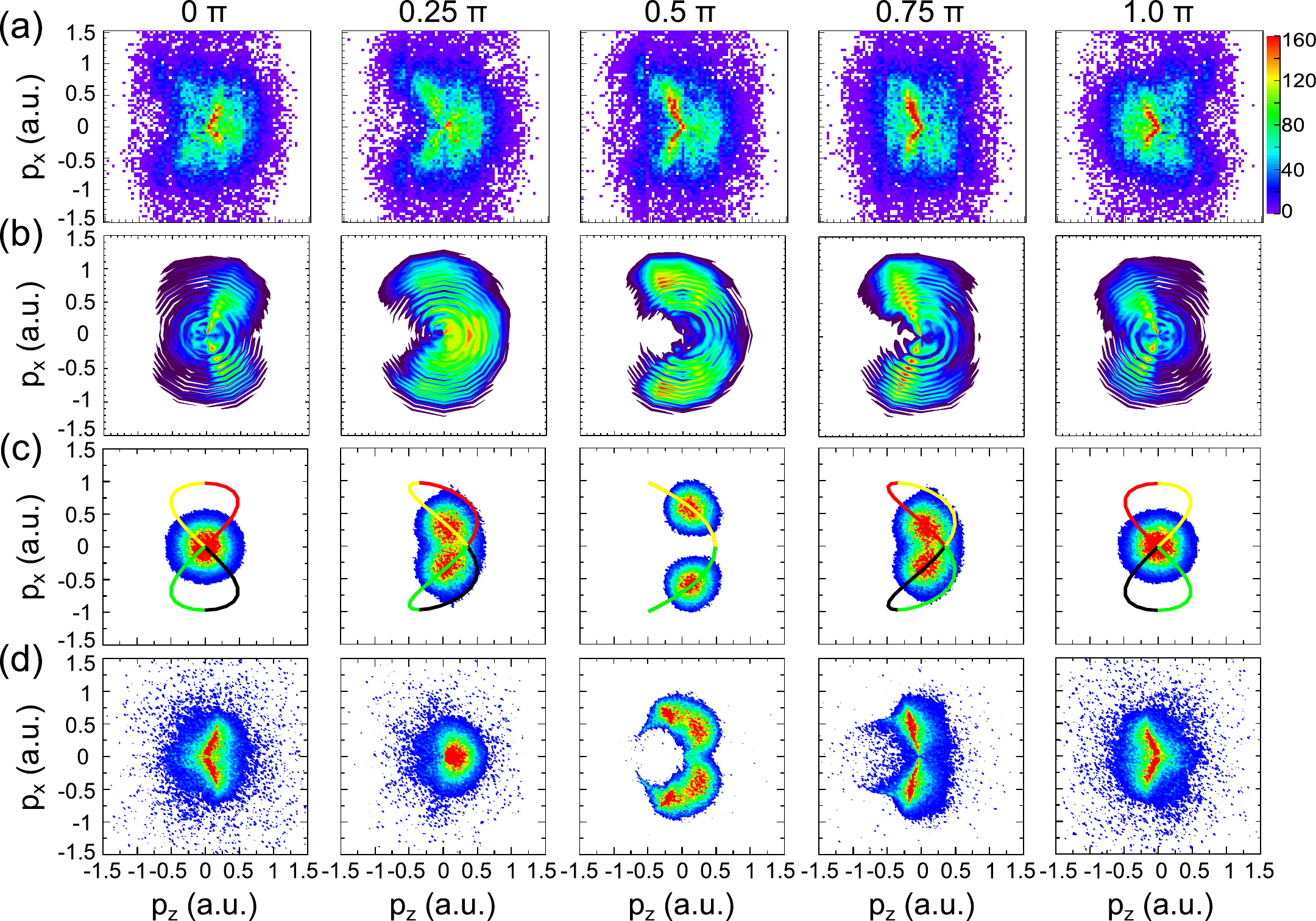}
\caption{Comparison of experimental (a) and calculated (b-d) electron momentum distributions. (b) Solutions of the TDSE. (c,d) CTMC simulations without (c) and with inclusion (d) of the ionic potential. See text for details.}
\label{fig3}
\end{figure*}

Measured electron momentum distributions in the polarization plane correlated to Ne$^+$ are shown in Fig.\,\ref{fig1}(c). The spectra are sensitive to the shape of the OTC field, i.e., to the relative phase $\Delta \varphi$. The electron spectra show a prominent x-shaped central structure and weaker fine-scale modulations. The latter are due to wave packet interferences \cite{Xie2012_interferometry}. 
The x-shaped central structure of the photoelectron spectra in Fig.\,\ref{fig1}(c) can be related to the 2D-evolution of the vector potential via $\vec{p}=-\vec{A}(t_0)$ that is also reflected in the SFCT spectra [cf. Fig.\,\ref{fig1}(b)]. However, this correspondence fails, e.g., for 
$\Delta \varphi=0$ and $\pi$, 
and there is also a disagreement between $\Delta \varphi=0.25\pi$ and $0.75\pi$ for which the classical relation $\vec{p}=-\vec{A}(t_0)$ predicts identical spectra, but the measured spectra are markedly different.
In the following we will show that in the presence of the parent ion the applicability of the sub-cycle mapping provided by this classical relation becomes limited, and we will provide a detailed analysis of the range of its applicability.

For an in-depth analysis of the photoelectron momentum distributions beyond the classical relation $\vec{p}=-\vec{A}(t_0)$ and the SFCT we performed both a three-dimensional TDSE and a classical trajectory Monte Carlo (CTMC) simulation. 
The TDSE was solved in velocity gauge for the Ne atom described in single active electron approximation using partial wave expansion, subjected to an OTC pulse, whose field is given by Eq.\,(\ref{OTC_field}), with $\omega=0.057$\,a.u. ($\lambda=800$\,nm), and $\hat{E}=0.0534$\,a.u., corresponding to an intensity of $10^{14}$\,W/cm$^2$ for the fields of both colors. The pulse envelopes were chosen as $f_x(t)=f_z(t)= \sin^2(\pi t/ T )$, with $T$ such that the duration is 6 cycles of $\omega$.
We neglect ionization from the Ne $2s$ orbital and consider only ionization from the $2p_m$ orbitals, with $m=0,\pm 1$. The electron spectra obtained are averaged over the initial $m$ sublevels. The effective potential is chosen to mimic the ionization energy of the $2p$ state to within 1\% of the literature value \cite{oep}. 
A comparison of the TDSE results with the measured electron momentum spectra is made in Fig.\,\ref{fig3}, 
where a common background has been removed from the experimental spectra in Fig.\,\ref{fig1}(c) by partial subtraction of a phase-integrated sum spectrum, similar as in \cite{Xie2012_interferometry}, and matching the color scale to that of the TDSE spectra.
In contrast to the SFCT the TDSE very well reproduces the experimental phase-dependent asymmetries.

%
\noindent 
\begin{figure}[tb]
\includegraphics[width=\columnwidth]{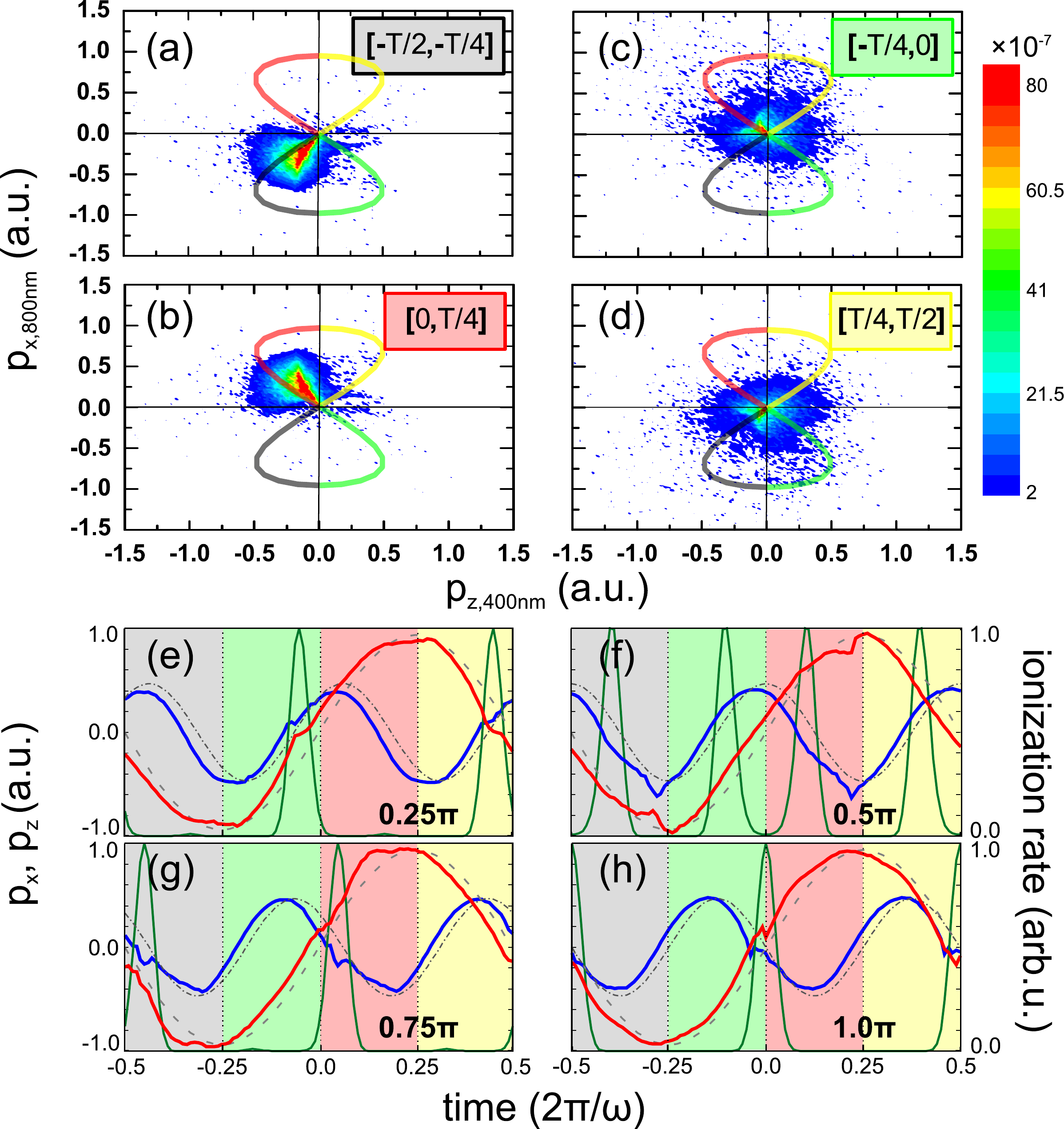}
\caption{(a)-(d) Electron momentum spectrum for $\Delta \varphi=\pi$ simulated by CTMC with the ionic potential included, separated into the contributions from trajectories born during different quarter-cycles of the OTC field, as indicated in the panels. (e)-(f) Average electron momentum $p_x(t_0)$ (red) and $p_z(t_0)$ (blue) over birth time $t_0$ calculated by CTMC, in comparison with $p_{x,z}=-A_{x,z}(t_0)$ (gray dashed and dotted-dashed lines, respectively) for various $\Delta \varphi$ as indicated in the panels. The green full lines denote ionization rates \cite{Yudin2001a} normalized to their respective maxima. Color coding of quarter cycles as in Fig.\,\ref{fig1}(a).}
\label{fig4}
\end{figure}

To further examine the origin of the discrepancy between experiment and the SFCT prediction, we now compare the experimental spectra to CTMC simulations with and without the influence of the parent ion potential on the electron trajectories. In brief, we solve Newton's equations of motion for an electron released from the $2p$ orbital of a neon atom interacting with an OTC laser field as given by Eq.\,\ref{OTC_field}. The electron release rate 
and the initial momentum distribution of the trajectories  perpendicular to the instantaneous OTC field vector are given by tunneling theory \cite{Delone1991}.
After ionization the electron is placed at the outer point where the field-suppressed Coulomb potential equals the binding potential, respectively at the origin when the Coulomb potential is turned off. Further details of the simulations can be found in, e.g., Refs. \cite{Chen2002, Wu2013_NSDI_elliptical}.
The parameters of the OTC field are $\omega = 0.057$\,a.u., $\hat{E} = 0.0534$\,a.u. (the same as in the TDSE simulations) and the pulse envelopes $f_x(t)$ and $f_z(t)$ are 1 for the first 18 and 36 cycles, respectively, and then gradually decrease to zero within three and six cycles, respectively, to imitate the experimental conditions.

Results of the CTMC simulations without the Coulomb field are shown in Fig.\,\ref{fig3}(c). The CTMC spectra resemble the SFCT spectra from Fig.\,\ref{fig1}(b), from which they differ by an accurate quantitative treatment of the initial electron momentum distribution \cite{Delone1991}. Because of that the spectral structures become wider and start to overlap, and therefore the x-shaped structures are somewhat blurred. We have checked that the SFCT results are recovered when the momentum spread is artificially limited to about 0.3\,a.u.
In order to understand the role of the ionic Coulomb potential in the photoelectron dynamics, we have performed another set of 
CTMC calculations in which the Coulomb potential $V(r)=-1/r$ has been included. 
The results of these simulations are shown in Fig.\,\ref{fig3}(d).
In comparing Figs.\,\ref{fig3}(c) and (d) it is striking how strongly the Coulomb field of the parent ion distorts the final electron momenta. 
The good qualitative agreement between the TDSE (b) and the CTMC (d) simulations verifies that the main features of the quantum calculation are captured very well in the classical picture. These results clearly show that the deviations of the measured photoelectron momentum distributions from the classical model (distortions and asymmetries) are caused by a purely classical effect, namely the additional driving force of the ionic Coulomb potential.

To identify those classes of trajectories that experience a strong distortion by the parent ion's Coulomb field we
separated the photoelectron momentum distribution into four parts based on a selection of trajectories in the CTMC simulations that are launched during different quarter cycles, as shown in Figs.\,\ref{fig4}(a)-(d) for $\Delta \varphi=\pi$. 
From these separated spectra it becomes clear that electrons emitted during quarter-cycles after the maximum of the $\omega$-field (black, red) [Figs.\,\ref{fig4}(a) and (b)] are closely mapped onto the field-dictated momentum value $\vec{p}=-\vec{A}(t_0)$, but experience Coulomb focusing by the ionic potential \cite{Brabec1996a}, as can be seen from their narrower momentum distributions as compared to the distributions without Coulomb potential [cf. Fig.\,\ref{fig3}(c)]. On the other hand, trajectories of electrons emitted before the $\omega$-field maximum (green, yellow) [Figs.\,\ref{fig4}(c) and (d)] are so strongly affected by the Coulomb field that their momenta are scattered over a wide range with a maximum close to zero momentum and, thus, the time-to-momentum mapping of the OTC field is lost.

In linearly polarized single-color laser fields, trajectories born during quarter-cycles after the field maximum recollide with the parent ion. Although in OTC fields the recollision condition needs to be fulfilled in two spatial dimensions for both colors, which is the case only for certain $\Delta \varphi$, our experiment suggests that -- on average -- direct trajectories are more strongly affected by the ionic potential than recolliding electrons, in agreement with recent results obtained for linearly polarized two-color pulses \cite{Xie2013}.
The surprisingly strong defocusing or scattering of the direct trajectories can be explained with a Coulomb field that counteracts the laser field and holds back the departing electrons, especially at the first few turning points of the direct trajectories where the parent ion's field has a strong influence.
For recollision trajectories in contrary, the Coulomb force at the first turning point, when the electron starts to return to the ion, is pulling in the same direction as the laser field. Thereby, the recollision velocity is increased, which renders the relative influence of the Coulomb driving force comparatively less important, leaving merely Coulomb focussing in the lateral direction.

In order to understand in detail the sub-cycle dynamics underlying the different behaviour of the direct and recolliding trajectories, we extracted from the CTMC spectra in Fig.\,\ref{fig3}(d) the final average electron momentum for each birth time, $\vec{p}(t_0)$, obtained by integration over the initial electron momentum distribution. The values $p_x(t_0)$ and $p_z(t_0)$ are depicted in Figs.\,\ref{fig4}(e)-(h) for different $\Delta \varphi$, in comparison with $p_{x,z}=-A_{x,z}(t_0)$ of purely field-driven trajectories assumed in the time-to-momentum mapping of OTC fields \cite{Kitzler2007, Kitzler2008, Shafir2009}. The comparison reveals that on the large scale the 
electron momentum $\vec{p}(t_0)$ roughly follows the value $-\vec{A}(t_0)$. However, there is a suspicious phase-shift between them, and for some $t_0$ there are also fast wiggles visible. Fig.\,\ref{fig1}(d) visualizes how these deviations translate into the overall electron momentum spectra. To calculate these spectra we have applied a procedure analogous to the SFCT described above, but summing up the values $\vec{p}(t_0)$ instead of $-\vec{A}(t_0)$. Note that only sub-cycle periods with an essential ionization rate in Figs.\,\ref{fig4}(e)-(h) contribute substantially to this procedure. 
The resulting spectra in Fig.\,\ref{fig1}(d) 
capture the experimentally observed asymmetries and distortions to a large degree. The overall phase-shift between $\vec{p}(t_0)$ and $-\vec{A}(t_0)$ is visible in  Fig.\,\ref{fig1}(d) as a shift of the transition between quarter-cycles from the pure field-induced mapping. 
This results in a timing failure of the mapping on the order of $2\pi/(32\omega)$. The fast wiggles, most visible for the (direct) green and yellow quarter-cycles, result in the severe spectral distortions such as the missing parts for $\Delta \varphi=0/\pi$.
%


In conclusion, we investigated the applicability of the sub-cycle time-to-momentum mapping provided by OTC fields with respect to the influence of the parent ion on the mapping.
We found that depending on their sub-cycle birth time the trajectories of photoelectrons are affected differently by the ion's Coulomb field.
While recollision trajectories are focussed, direct trajectories are defocussed or strongly scattered. 
The observed Coulomb defocussing of direct trajectories adds to the variety of Coulomb distortions and likely accounts for the large angle scattering part of the Coulomb distorted transverse photoelectron momentum in linear polarization \cite{Comtois2005}.

We acknowledge funding by the Austrian Science Fund (FWF) under grants P21463-N22, P25615-N27, SFB-F49 NEXTlite, by a starting grant from the ERC (project CyFi), by the National Basic Research Program of China (grants 2013CB922201 and 2011CB808102) and the NNSF of China (grant 11334009), and by the Australian Research Council in the form of the Discovery grant DP120101805. Resources of the Australian National Computational Infrastructure (NCI) Facility were employed.


\clearpage

\setlength{\textwidth}{16.5cm}     
\setlength{\oddsidemargin}{0cm}   
\setlength{\evensidemargin}{0cm}  
\setlength{\textheight}{22.5cm}   
\setlength{\topmargin}{0cm}       
\setlength{\headheight}{0cm}      
\setlength{\headsep}{0cm}         
\setlength{\footskip}{1cm}       

\renewcommand{\thefigure}{\arabic{figure}-SM}

\onecolumngrid

\begin{center}
{\bf \sffamily \huge \color{somtitlecolor} Supplemental Material}
\end{center}

\section{Calibration of the relative phase $\Delta \varphi$}

\begin{figure}[b]
\includegraphics[width=0.7\columnwidth]{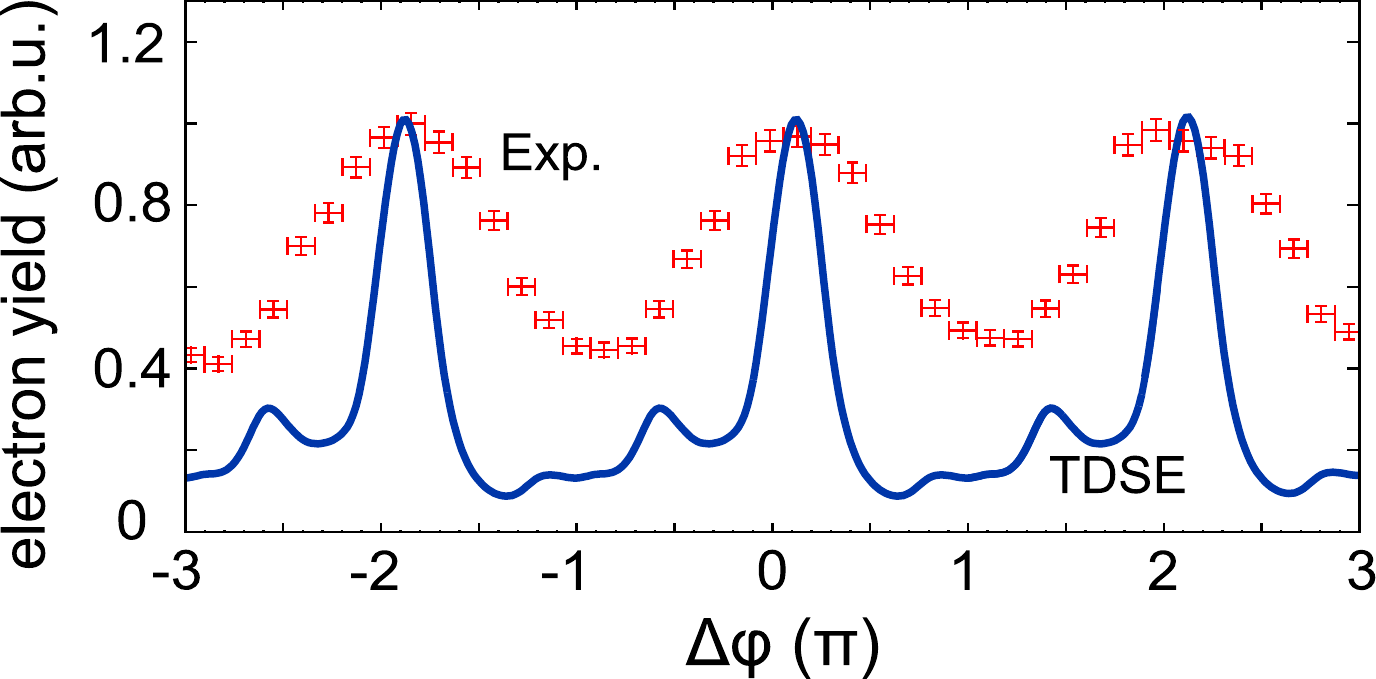}
\caption{Calibration of $\Delta \varphi$ based on a comparison of the yield of electrons correlated with Ne$^+$ in the spectral cut-off region $p_z > 1.5$\,a.u. (other directions integrated over) between experiment (data points) and solutions of the TDSE (solid line).} \label{fig1som}
\end{figure}

The interpretation of the experiment described in the main document 
depends on the calibration of the phase relation between the fundamental and the second harmonic beam. We use the Ne$^+$ yield in the spectral cut-off region $p_z > 1.5$\,a.u. for calibration, since there distortions in the momentum due to Coulomb effects are small. We calculate the ionization probability of Ne by solving the two-dimensional time-dependent Schr\"odinger equation (TDSE) for a model atom \cite{Xie2012_interferometry, Xie2013}. In Fig.\,\ref{fig1som} we plot the calculated probability and the measured yield over $\Delta \varphi$.

\noindent 

Because the maxima of the vectorial laser electric field are largest at $\Delta \varphi=n\pi$, $n \in \mathbb Z$, tunneling theory \cite{Yudin2001} predicts highest ionization yield for these relative phases. Because electron emission takes place asymmetrically along $z$ in OTC fields, see e.g. Fig.\,1(c) in the main document, electron emission peaks only every $2\pi$ into the hemisphere $p_z>0$.
Accordingly, the results of the TDSE simulation in Fig.\,\ref{fig1som} show maxima of the yield at $\Delta \varphi \approx n2\pi$, $n \in \mathbb Z$. 
A detailed inspection of the simulated yield modulation shows that there is a slight shift of the maxima away from the values $\Delta \varphi = n2\pi$, $n \in \mathbb Z$.
This shift was also observed in Ref.\,\cite{Xie2013} for parallel polarization directions of the two colors.
As the ionization process is a highly dynamical process, many effects such as electronic excitation and interference during ionization or Coulomb trapping of outgoing and/or returning electron wave packets can be the reason for this shift. 
Furthermore, as these dynamic processes depend on the laser parameters, in particular on intensity, this small shift may also be intensity dependent. Ultimately this shift leads to a small uncertainty in the precision of our phase-calibration. 
For parallel polarization directions it was shown by two-dimensional and three-dimensional solutions of the TDSE for different shapes of the binding potential that the shift is on the order of $|0.1\pi|$ over a large range of laser pulse parameters \cite{Xie2013}. The uncertainty in calibrating $\Delta \varphi$ by the here described method can therefore be estimated to be $\pm 0.1\pi$.

For all presented experimental data in the main document we define the relative phase $\Delta \varphi$ so that the maxima of the measured photoelectron yield (red data points in Fig.\,\ref{fig1som}) correspond to the maxima of the calculated ionization probability (blue line in Fig.\,\ref{fig1som}). 
The sign of the phase was determined from the asymmetry of the measured electron momentum spectrum at $\Delta \varphi = 0.5\pi$ in comparison to the spectrum obtained by solving the time-dependent Schr\"odinger equation, see Fig.\,2 in the main document. This phase calibration was used throughout the paper.


\vspace{1cm}

\twocolumngrid

\end{document}